\documentclass[]{aa}
\usepackage{graphicx}
\begin{document}
\title{ABELL 43, a second pulsating ``hybrid-PG~1159" star\thanks{Based on  observations made with  the 
Nordic Optical Telescope, operated on the island of La Palma jointly by Denmark, Finland, Iceland, Norway, 
and Sweden, in the Spanish Observatorio del Roque de los Muchachos of the Instituto de Astrofisica de Canarias. }}
\author{G. Vauclair\inst{1}
 \and J.-E. Solheim\inst{2}
 \and R.H. \O stensen\inst{3}}
\offprints{G. Vauclair}
\institute{Universit\'e Paul-Sabatier, Observatoire Midi-Pyr\'en\'ees, CNRS/UMR5572,
 14 Av. E. Belin, 31400 Toulouse, France
 \and Institute of Theoretical Astrophysics, University of Oslo, p. Box 1029,
 N-0315, Blindern, Oslo, Norway
 \and Isaac Newton Group, Apartado de correos 321, E-38700, Santa Cruz de la Palma, Spain}
\titlerunning{The Abell 43 pulsator}
\date{Received;accepted}
\maketitle
\begin{abstract}
We report   observations of the planetary nebula nucleus Abell 43, obtained at the 2.5m Nordic Optical Telescope, 
which show that it is a pulsator. Abell 43, a ``hybrid-PG 1159'' type star, is the second pulsator of this class, after 
HS 2324+3944. From the limited data set acquired, we find that Abell 43 exhibits at least two periods of 2600~s and 
3035~s, the longest ones observed up to now in PG 1159 and ``hybrid-PG 1159'' pulsators. This strongly 
suggests that the variations are due to non-radial g-mode pulsations and cannot be
a consequence of binarity. This discovery raises puzzling questions regarding the excitation mechanism in this 
H rich,  C and O  poor ``hybrid-PG 1159'' since the C and O abundances are too low to trigger the instability through the 
$\kappa$-mechanism induced by the partial ionization of C and O, a mechanism 
invoked to explain the instability in the PG 1159 stars and in the previously known ``hybrid-PG 1159''
pulsator HS 2324+3944. 
\end{abstract}

\section{Introduction}

The instability mechanism for the pulsating PG 1159 stars (or GW Vir stars), named after the prototype of this 
class PG 1159-035, has been
 identified as the $\kappa$-mechanism related to the partially ionized carbon and oxygen 
by Starrfield et al. (\cite{star84}, \cite{star85}). While it was subsequently claimed that a small admixture
of hydrogen would inhibit the instability (Stanghellini et al. \cite{stan91}), Saio (\cite{saio96}) found that
the stability of the g-modes was not affected by a 3\% admixture of hydrogen by mass in the 
otherwhise characteristic composition of PG 1159 stars. In the mean time, Napiwotzki \&
Sch\"onberner (\cite{napi91}) discovered a new class of PG 1159 stars showing strong Balmer lines,
which they called ``hybrid-PG 1159" stars.  Werner (\cite{wern92}) classified  them as lgEH 
in his classification scheme.

Silvotti (\cite{silv96}) found the first pulsating ``hybrid-PG 1159" star: HS 2324+3944. The pulsation 
spectrum was dominated by a 2140~s period. This was later  confirmed  by
Silvotti et al. (\cite{silv99}) who found HS 2324+3944 to be a rich multiperiodic pulsator with at 
least 7 frequencies between 391~$\mu$Hz (2553~s) and 961~$\mu$Hz (1039~s). 
While Bradley \& Dziembowski (\cite{brad96}) and Cox (\cite{cox03}) require a different composition in the 
driving zone, compared to the composition observed at the surface, Gautschy (\cite{gaut97}) was
able to reproduce the instability strip for the PG 1159 pulsators and the observed range of
periods for the then unique ``hybrid-PG 1159" HS 2324+3944, with a uniform composition in 
agreement with the observed composition. More recently, Quirion et al. (\cite{quir04}), reanalyzing
the stability problem of the PG 1159 stars, confirm that the instability strip of the 
PG 1159 stars is well reproduced by models having a uniform composition similar to the 
observed one for each PG 1159 pulsator. They also 
reproduce satisfactorily the periods observed in HS 2324+3944, with a model including
10\% of hydrogen mass fraction,  and find unstable modes 
in a period range 1355~s-3013~s, slightly wider than  the observed one.
They also show that small differences in He/C/O composition and metallicity
from star to star account 
for the puzzling coexistence of pulsators and non-pulsators with identical atmospheric 
parameters.

In this context, it is worthwhile to wonder whether HS 2324+3944 is a unique case of 
pulsating
``hybrid-PG 1159" star, and if more such stars do exist, were are the boundaries of the
corresponding instability strip. Ciardullo \& Bond (\cite{ciar96}) conducted a large survey to search
for pulsating nuclei of planetary nebulae. Included in this survey were 3 of the ``hybrid-
PG 1159" stars known by the time: Abell 43, NGC 7094 and Sh 2-68. None of them were found 
to pulsate and their table 2 lists them as non-pulsators. However, in the case of Abell 43,
they noted that the FT of their light curve did show one peak at $\approx$404~$\mu$Hz with an
amplitude of 3.5~mmag, which was just below their  adopted detection limit.  For this reason, 
they did not consider Abell 43 as a pulsator. In the case of NGC 7094, they did not detect any significant
peak in the power spectrum above 300~$\mu$Hz, but they comment  that their light curve shows quasi-sinusoidal
variations on a time scale of 2~hr and with an amplitude of 20~mmag. 
More recently, Gonz\'alez P\'erez (\cite{gonz04}) reobserved both Abell 43 and NGC 7094 
and did not detect any variations. For Abell 43, the maximum
amplitude of its power spectrum was $\leq$3 mma (where mma, the milli-modulation amplitude,
 is the fractional amplitude in the Fourier
Transform in units of $10^{-3}$),  while it would have required an amplitude of 6 mma
for a peak in the power spectrum to have a False Alarm Probability (FAP) of only 1/20 to be due to noise
 and be considered as possibly significant. In the case of  NGC 7094, for wich he obtained better quality data, the maximum amplitude in the 
power spectrum was $\leq$0.7 mma  while for the same FAP  it would have required an amplitude of 0.9 mma
to be significant at the same level of confidence.
However, from his data set on NGC 7094, one cannot check whether the 2~hr time scale variability suspected by
Ciardullo \& Bond (\cite{ciar96}) is present since the run is shorter.

In the present letter, we report new observations of Abell 43 showing that it is indeed
a pulsator, the second one among the class of the ``hybrid-PG 1159" stars. In section 2 we
describe our observations. We discuss the results in the context of the instability 
mechanism in section 3.

\section{Observations and Analysis}

We used the Nordic Optical Telescope (NOT) with the Andalucia Faint Object Spectrograph and Camera
(ALFOSC)  equipped with a  thinned 2048 $\times$ 2048 E2V CCD 42-40  chip, operated in multi-windowed fast photometry mode
 (\O stensen \cite{oste00}). 
We used a filter centered on 550~nm with a FWHM of 275~nm, in order to increase the contrast 
with the sky background.
Flat
field measurements were performed at the beginning of each observing night, and the bias level
was determined from overscan obtained for each windowed frame.
The light curves have been corrected for atmospheric extinction by using a coefficient proportional to the 
airmass. The coefficient is adjusted so as to flatten the light curve as much as possible. 

\begin{figure}
\resizebox{\hsize}{!}{\includegraphics{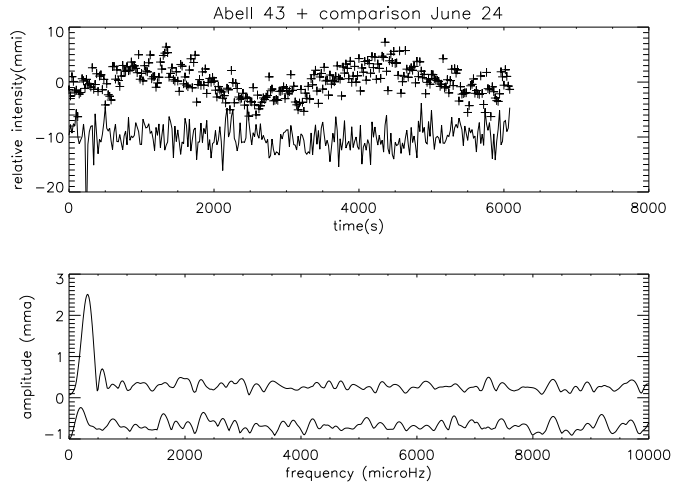}}
\vspace*{0.5cm}
\caption{The upper panel shows the normalized light curve of Abell 43 obtained on June 24th, 2004. 
The relative intensity in units of milli-modulation intensity (mmi), which is the fractional modulation intensity 
in units of $10^{-3}$,  is plotted as a function of time in seconds.
The   Abell 43 light curve (crosses) is shown together with  the average light curve for one of the comparison stars (full line)
relative to the average of the remaining comparison stars. The comparison star light curve is displaced 10 mmi downwards for clarity.
The lower panel shows the FT of these light curves, plotted  in units of milli-modulation amplitude (mma)  as a 
function of frequency in $\mu$Hz. The FT for the comparison star light curve is displaced downwards by 1 mma for clarity.}
\label{fig1}
\end{figure}

\begin{figure}
\resizebox{\hsize}{!}{\includegraphics{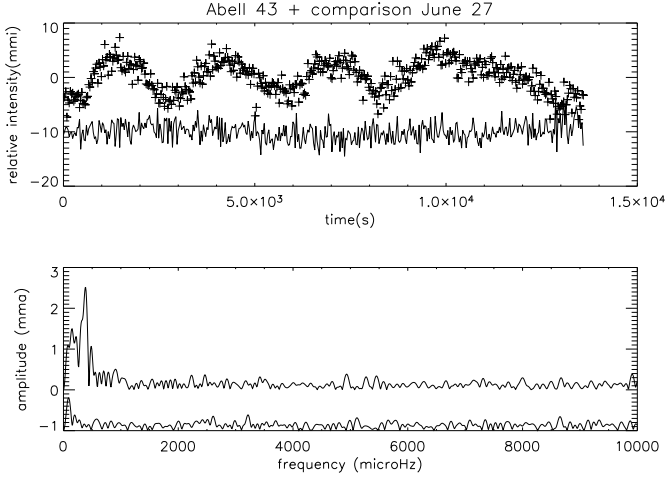}}
\vspace*{0.5cm}
\caption{As in Fig.~\ref{fig1} for the run on June 27th, 2004.}
\label{fig2}
\end{figure}

We first observed Abell 43 on June 24th, 2004 (start time 22:25:28 UT) together with 6 comparison stars and 2 sky
background fields. The observing cycle was choosen as 20~s, which, with a total readout 
time of  7.4~s
for the 9 windows, results in an exposure time of  12.2~s for each frame. The average seeing was 0.9 arcsecond (FWHM) 
so that  the  S/N ratio was optimized with a 12 pixels aperture, which corresponds to 
 an aperture diameter of 4.6 arcseconds at the ALFOSC pixel scale (0.19 arcsecond/pixel).
 We obtained 
a 6000~s light curve, showing unambiguously that the star is variable. The light curve 
of the target and one of the comparison stars is 
shown in Fig.~\ref{fig1}, together with their   Fourier transform, plotted in amplitude terms.
 Two cycles are clearly visible
in the light curve. The FT shows an unresolved peak of 2.5~mma amplitude, while the average 
 noise level is about 0.3~mma. 
We observed  Abell 43 again on June 27th, 2004 (start time 23:15:26 UT) for a confirmation run. 
We used 3 comparison stars and 2 sky background fields. The  cycle time was 30~s, which, with a total readout time of 5~s for the 6
windows, results in an exposure time of 24.7~s for each frame.
The average seeing was 1.5~arcsecond (FWHM). The best S/N ratio was obtained with an aperture of 17 pixels, i.e. 6.5~arcseconds.
We obtained a longer light curve of  ~13500~s, shown in Fig.\ref{fig2}, together with its Fourier transform. 
On this light curve, one clearly sees that the variation cycles are irregular in length, suggesting that the star is 
multiperiodic or that a frequency change occured during the run.  
 The average noise 
level in the FT is $\sigma$=~0.2~mma. The peak at 340~$\mu$Hz with an amplitude of 2.5~mma, is at
 12.5~$\sigma$.

\begin{figure}
\resizebox{\hsize}{!}{\includegraphics{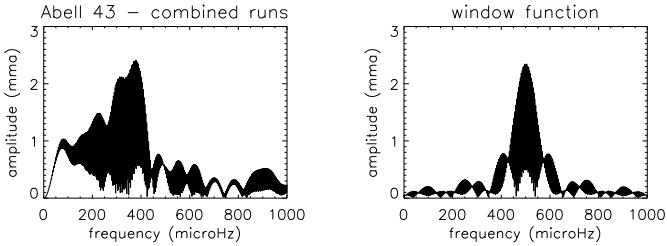}}
\vspace*{0.5cm}
\caption{The left  panel shows the FT of the combined light curves (June 24th + 27th).
 The amplitude in units of milli-modulation amplitude (mma) is plotted as a 
function of frequency in $\mu$Hz. The FT is shown on the restricted frequency range 0-1000~$\mu$Hz. The right panel shows 
the window function on the same scale.  }
\label{fig3}
\end{figure}

In spite of the fact that these  two observing runs are three days apart, we combined the two
light curves to perform a FT with a better resolution, shown in Fig.~\ref{fig3} together with the corresponding
window function. Both are shown on an enlarged scale encompassing the frequency interval
which shows power, i.e. betwen 0 and 1000 $\mu$Hz. The FT  shows  at least two significant 
peaks at 329~$\mu$Hz  (3035~s period), with a 2.1~mma amplitude, 
 and at 384~$\mu$Hz (2600~s period), with a 2.4~mma amplitude.

\begin{figure}
\resizebox{\hsize}{!}{\includegraphics{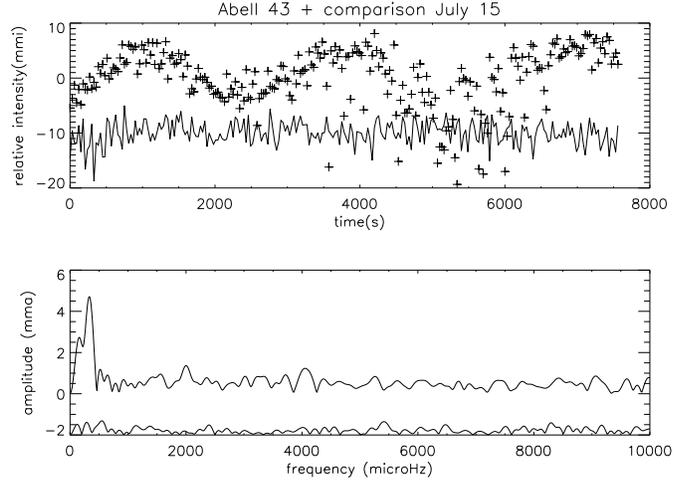}}
\vspace*{0.5cm}
\caption{As in Fig.~\ref{fig1} for the run  on July 15th, 2004. 
 The FT for the comparison star light curve is displaced downwards by 2 mma for clarity.
Comparison with Fig.~\ref{fig1} and Fig.~\ref{fig2} clearly shows the amplitude variations.}
\label{fig4}
\end{figure}

A third run was obtained on July 15th, 2004 in similar conditions than the two previous ones, except that the aperture was 10 pixels,
i.e. 3.8~arcseconds, and the cycle time was 30~s with an integration time of 25~s. The light curve, $\approx$7500~s long, 
 and its FT are shown in Fig.\ref{fig4}. This third run shows 
variations with an amplitude significantly larger since the peak in the FT is at 4.7~mma.

Previous observations did not detect the Abell 43 variations because  1) they have long periods and 2) their amplitudes are small and  variable.

\section{Discussion}

 Are the variations observed in Abell 43 due to pulsations? A number of planetary nebulae nuclei are
binary sytems. Some reflection effects may happen if the companion is a hot star close enough to 
the planetary nebula nucleus and/or the nucleus itself may take a non spheroidal shape because of the 
tidal interaction induced by the companion. Both effects would produce a light variation in phase with 
the rotation period of the nucleus.
 In a search for companions of planetary nebula nuclei, Ciardullo et al. (\cite{ciar99})
 did not find any evidence
for such a companion close to Abell 43 nor to NGC 7094. However, their method would have detected only visual companions, too 
far from the planetary nebula nucleus to produce any reflection effect. This does not preclude the possibility
for Abell 43 and NGC 7094 to have a close companion. Such a close binary system would be detectable  through radial
velocity measurements. We are not aware of any such  radial velocity study on those two stars.
 But the multiperiodicity or the frequency variation 
found in the FT of the Abell 43 light curve excludes that the observed variations could be due to such an effect
in that star.
We do conclude that the  variations in Abell 43 are due to non-radial $g$-mode pulsations. With periods of 2600~s and 3000~s,
Abell 43 shows the longest periods observed in PG 1159 and ``hybrid-PG 1159" stars.
In the case of NGC 7094, we clearly need more data to check the reality of the  2~hr variability
suspected by Ciardullo \& Bond (\cite{ciar96}).

 The discovery of pulsations in Abell 43 is puzzling and raises  interesting questions regarding 
the excitation mechanism and the evolutionay status.
Both Abell 43 and NGC 7094 occupy the same location in the log~$g$- $T_{eff}$ diagram (Dreizler et al. \cite{drei95}). In this
diagram, they seem to lie on the same evolutionary track as HS 2324+3944, between for instance 
the evolutionary tracks
of Sch\"onberner (\cite{scho83}) for 0.565$M_{\odot}$ and 0.605$M_{\odot}$. With
$T_{eff}$= 110~kK and log~$g$= 5.7, they are clearly less evolved than HS 2324+3944 at $T_{eff}$= 130~kK and log~$g$=
6.2 (Dreizler et al. \cite{drei96}). Both Abell 43 and NGC 7094  have still a surrounding nebula while none is seen around HS 2324+3944.
 Most interestingly, their composition
strongly differs from HS 2324+3944 and PG 1159 composition, being H-rich (42\% by mass), with similarly a high
abundance of He (51\% by mass), and poor in C (5\% by mass) while N and O are below detection limit
(Dreizler et al. \cite{drei95}).
By comparison, HS 2324+3944 has the same relative C/He abundance as the PG 1159 stars but a much smaller H abundance 
 of 18\% by mass, with 37\% of He, 44\% of C and less than 1\% of N and O, by mass (Dreizler et al. \cite{drei95}).
A mass-loss rate has been measured in NGC 7094, of log$\dot{M}$= -7.3~$M_{\odot}yr^{-1}$ from CIV line or of 
log$\dot{M}$= -7.7~$M_{\odot}yr^{-1}$ from OVI line (Koesterke et al. \cite{koes98a}, \cite{koes98b}).
While no such mass-loss rate determination exists for Abell 43, Dreizler (\cite{drei98}) reports that both Abell 43
and NGC 7094 show a P Cygni profile in the O V line at 1371 \AA. This strongly suggests that Abell 43 is also losing mass.  
This  gives a coherent picture of an evolutionary link between Abell 43, NGC 7094 and HS 2324+3944, consistent with the fact that
HS 2324+3944 has less hydrogen left than Abell 43 and NGC 7094 as a consequence of the mass-loss, and with the fact that HS 2324+3944
had the time to disperse its nebula while Abell 43 and NGC 7094 kept their nebulae.
Finally, among those two stars,
Abell 43 is a pulsator while for NGC 7094 we can only say that we do not know yet.

The model explaining the instability in the  PG 1159 stars  and in the ``hybrid-PG 1159" HS 2324+3944 does not seem applicable
 to Abell 43. Both Abell 43 and NGC 7094 have a very low abundance of C and O. 
Even if one considers that the preliminary abundance analysis of Dreizler et al. (\cite{drei95}) does not take into account
the influence of the stellar wind on the atmospheric modeling, it is difficult to imagine that taking this effect  into account
could  increase  the C and O abundances so much as to reach  the values derived for HS 2324+3944. 
If the  composition of Abell 43 is uniform from the surface to the potential driving zone, as expected if mass-loss
precludes any gravitational settling, it is difficult to invoke an instability due to the $\kappa$-mechanism
induced by the C and O partial ionization with such a low abundance of C and O. As in the case of the PG 1159 pulsators,
one may have  also to explain the coexistence of pulsators  and non pulsators with identical atmospheric parameters among the 
``hybrid-PG 1159''stars if NGC 7094 is confirmed to be a non pulsator.

A first alternative would be to postulate that the still unknown  mass-loss rate in Abell 43 is too low to  inhibit
 the gravitational
settling of C and O. The composition in the driving zone would not reflect the abundances observed at the surface.
The C and O abundances in the driving zone could be enhanced such as to trigger the pulsations through the $\kappa$-mechanism. However, if 
the gravitational settling takes place, it should also affect He relative to H abundance. The observed  He
and H abundance in Abell 43 does not fit with this scheme. In addition, it is expected that the mass-loss rate
in Abell 43 should be higher than in HS 2324+3944, or at least similar to the one measured in NGC 7094, 
if it is related to the luminosity. If this is the case, the gravitational
settling should not occur in Abell 43 if it does not occur in HS 2324+3944. Furthermore, if there is no mass loss
in Abell 43 there cannot be any evolutionary link with stars like HS 2324+3944 and PG 1159 stars which  
are hydrogen deficient. 

Another possibility to trigger the instability, through the $\epsilon$-mechanism induced by remnant nuclear shell burning during the planetary nebula
and pre-white dwarf evolutionary phases, was suggested by Kawaler et al. (\cite{kawa86}). One may wonder whether the Abell 43 variability
could be due to this mechanism. However, it would make only the low order $g$-modes unstable, which have periods much shorter than the observed 
periods in Abell 43. None of the surveys of planetary nebulae nuclei conducted by Grauer et al. (\cite{grau87}), Hine \& Nather (\cite{hine87}) 
and more recently by Ciardullo \& Bond (\cite{ciar96}) found any short period variability which could be identified with $g$-modes triggered
 by the $\epsilon$-mechanism. The same conclusion was reached by Vauclair et al. (\cite{vauc02}) in their analysis of the PG 1159 central star
of the planetary nebula RX~J2117+3412.

A third alternative would be to give up the idea that the pulsations in Abell 43 are due to the $\kappa$-mechanism
related to the partial ionization of C and O or to the $\epsilon$-mechanism. We suggest that  episodic mass loss could excite the pulsations 
(by stochastic excitation). The same mechanism could also be at work in NGC 7094, where a mass-loss is observed and 
measured. This model predicts that NGC 7094 could also pulsate. 
In this context, it is worth pointing out the sudden change in the pulse length observed in Abell 43 during the June 27th run
 (see Fig.~\ref{fig2}). Similar
changes on a short timescale
have also been detected in the other pulsating planetary nebula nucleus NGC 246 (Gonz\'alez P\'erez \cite{gonz04}). Such
behaviours cannot be related to beating phenomena. They are real changes on short time scales which deserve longer observations to be investigated. 
This suggestion could be tested by observations. It would require simultaneous spectroscopic and photometric
observations to check whether 1) there is evidence of time dependent mass-loss in Abell 43 and 2) there is a correlation between
the pulsation amplitudes and the mass-loss episodes. Alternatively, the life time of the pulsation modes, if excited
by episodic mass-loss events, should be short compared to the life time of modes excited by the $\kappa$-mechanism.
This could  be checked by measuring the line width in the power spectrum. Long enough multisite campaigns may provide
the necessary frequency resolution to check the life time of the  modes. A wavelet analysis may be 
more appropriate to reveal those modes whose lifetime could be  short relative to the length of the observing campains.
 It is also urgently needed
to check whether NGC 7094 could  also be a pulsator, possibly with longer periods than the one which could have been determined from 
previous observations.   

\begin{acknowledgements}

The data presented here have been taken using ALFOSC, which is owned by the Instituto de Astrofisica de Andalucia (IAA) and
operated at the Nordic Optical Telescope under agreement between IAA and the NBIfAFG of the Astronomical Observatory of Copenhagen. 
We thanks P. Muhli for acquiring  data for us on July 15th, 2004.
\end{acknowledgements}


\begin{thebibliography}{}

\bibitem[1996]{brad96} Bradley, P. A. \& Dziembowski, W. A. 1996, ApJ, 462, 376
\bibitem[1996]{ciar96} Ciardullo, R. \& Bond, H. E. 1996, AJ, 111, 2332
\bibitem[1999]{ciar99} Ciardullo, R., Bond, H. E., Sipior, M. S., Fullton, L. K., Zhang, C.-Y., \& Schaefer, K. G. 1999,
AJ, 118, 488
\bibitem[2003]{cox03} Cox, A. N. 2003, ApJ, 585, 975
\bibitem[1998]{drei98} Dreizler, S. 1998, Baltic Astronomy, 7, 71
\bibitem[1995]{drei95} Dreizler, S., Werner, K., \& Heber, U. 1995, in White Dwarfs,Lecture Notes in Physics 443, 
eds.: D. Koester \& K. Werner, (Springer-Verlag: Heidelberg), 160 
\bibitem[1996]{drei96} Dreizler, S., Werner, K., Heber, U., \& Engels, D. 1996, A\&A, 309, 820
\bibitem[1997]{gaut97} Gautschy, A. 1997, A\&A, 320, 811
\bibitem[2004]{gonz04} Gonz\'alez P\'erez, J. M. 2004, PhD Thesis, University of Troms\o
\bibitem[1987]{grau87} Grauer, A. D., Bond, H. E., Liebert, J., Fleming, T. A., \& Green, R. F. 1987, ApJ, 323, 271
\bibitem[1987]{hine87} Hine, B. P. \& Nather, R. E. 1987, in The Second Conference on Faint Blue Stars, IAU Colloq. 95, eds. 
A. G. D. Philip, D. S. Hayes \& J. Liebert (Schenectady, NY: L. Davis Press), 619 
\bibitem[1986]{kawa86} Kawaler, S. D., Winget, D. E., Hansen, C. J., \& Iben, I. Jr. 1986, ApJ, 306, L41
\bibitem[1998a]{koes98a} Koesterke, L., Dreizler, S., \& Rauch, T. 1998a, A\&A, 330, 1041
\bibitem[1998b]{koes98b} Koesterke, L. \& Werner, K. 1998b, ApJ, 500, L55
\bibitem[1991]{napi91} Napiwotzki, R. \&  Sch\"onberner, D. 1991, A\&A, 249, L16
\bibitem[2000]{oste00} \O stensen, R. H. 2000, PhD Thesis, University of Troms\o
\bibitem[2004]{quir04} Quirion, P.-O., Fontaine, G., \& Brassard, P. 2004, ApJ, 610, 436
\bibitem[1996]{saio96} Saio, H. 1996, in Hydrogen-Deficient Stars, ASP Conf. Series 96, 
eds.: U. Heber \& C. S. Jeffery, 361
\bibitem[1983]{scho83} Sch\"onberner, D. 1983, ApJ, 272, 708
\bibitem[1996]{silv96} Silvotti, R. 1996, A\&A, 309, L23
\bibitem[1999]{silv99} Silvotti, R., Dreizler, S., Handler, G., \& Jiang, X.J. 1999, A\&A, 342, 745
\bibitem[1991]{stan91} Stanghellini, L., Cox, A. N., \& Starrfield, S. 1991, ApJ, 383, 766
\bibitem[1984]{star84} Starrfield, S., Cox, A. N., Kidman, R. B., \& Pesnell, W. D. 1984, ApJ, 281, 800
\bibitem[1985]{star85} Starrfield, S., Cox, A. N., Kidman, R. B., \& Pesnell, W. D. 1985, ApJ, 293, L23
\bibitem[2002]{vauc02} Vauclair, G., Moskalik, P., Pfeiffer, B. et al. 2002, A\&A, 381, 122
\bibitem[1992]{wern92} Werner, K. 1992, in Atmospheres of Early-Type stars, eds.: U. Heber \& C.S. Jeffery,
Lecture Notes in Physics 401, (Springer-Verlag: Heidelberg), 273

\end{thebibliography}
\end{document}